\def\aa{{A\&A}}

\def\aj{{AJ}}

\def\apj{{ApJ}}

\def\mnras{{MNRAS}}

\documentclass[cup5b]{caps}
\usepackage{graphicx}
\usepackage{amssymb}
\usepackage{ociwsymp1e}
\HeadText{E. Emsellem}

\begin{document}

\pagenumbering{arabic}

\author[]{E. EMSELLEM
\\ Centre Astronomique de Recherche de Lyon, Saint-Genis Laval, France }

\chapter{The Role of Inner Density Waves \\ in Fueling Galactic Nuclei}

\begin{abstract}
I present here a heterogeneous selection of galaxies in which inner
density waves (bars, spirals, $m=1$ modes) have been observed, and
which are presumed to host a supermassive black hole.
\end{abstract}

\section{Introduction}

Density waves the presence of which can be witnessed in structures such as
bars, spirals and lopsided distributions are ubiquitous in disk
galaxies. If we believe that central massive dark objects are present
in most nearby galaxies, then density waves and black holes should
often be concomitant.  This paper is intended as an attempt to
illustrate this issue by providing a few examples of inner ($R <
		1$~kpc) density
waves in galaxies which are presumed to host a supermassive black
hole. We have found that inner density waves are often easily traceable
via a combined simple morphological and dynamical analysis, but often
ignored in the subsequent dynamical modeling.  We also briefly mention
some evidence for gas fueling in the central few tens of parsecs via
density waves.

\section{The Prototype Seyfert~2: NGC~1068}

One of the best evidence for gas streaming towards the centre of a
nearby galaxy hosting a central dark mass is given by the well-studied
Seyfert~2 galaxy NGC~1068. Indeed, a near-infrared bar is well visible
in the central 20 arcseconds of this spiral galaxy, and streaming
motions have been uncovered in the gas kinematics (see e.g. Schinnerer
		et al. 2000 and references therein). Note that an outer
oval structure has been advocated by Schinnerer et al. to be a
large-scale (weak) primary bar, implying that the inner bar is in fact
a secondary bar. Although this is one of the best studies active
galactic nucleus, it is surprising to realize how little we
know about the gravitational potential in the central few kpc of this galaxy.

We have recently obtained integral field spectrography of this inner
region using {\sc SAURON} at the WHT. We first had to carefully
disentangle the respective contributions of the ionized gas and of the stellar
component. This was achieved by performing an optimal stellar template
fit using an iterative procedure and a library of synthetic stellar
spectra kindly provided by Alexandre Vazdekis (see Vazdekis 1999 and
references therein).

This allowed us to probe both the stellar and gaseous kinematics with
unprecedented field coverage. The signature of the so-called near-infrared
bar is clearly observed in the {\sc SAURON} stellar velocity map in
the form of an ''S-shaped'' isovelocity curve.  The ionized gas
distribution and kinematics is complex, particularly in the central 5
arcseconds or so, where the influence of the AGN is dominant,
with the presence of e.g. strong outflows and scattered light. We can
however still trace the gas streaming on the leading edge of the inner
bar, along spiral structures directed towards the outer two-arms
spiral observed in the CO maps (Schinnerer et al. 2000).
\begin{figure}
\centering
\includegraphics[width=110mm]{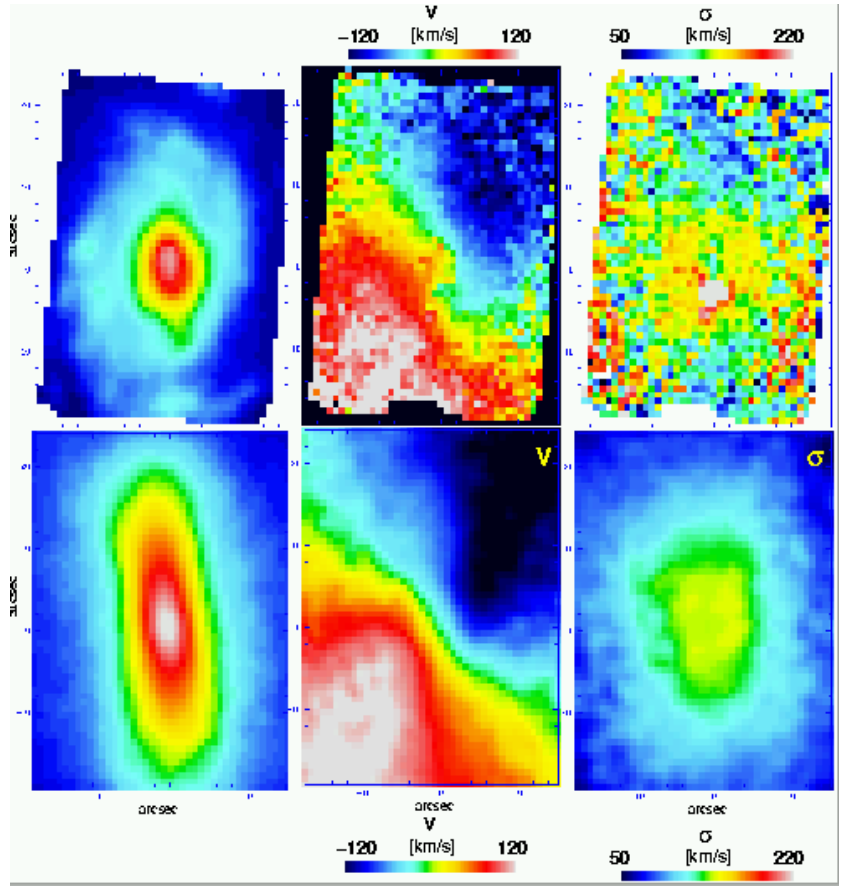}
\caption{Two-dimensional stellar kinematics of NGC~1068 as observed
with the {\sc SAURON} spectrograph (top panels), and from a numerical N-body
$+$ SPH simulations (bottom panels, see text). From left to right:
surface brightness reconstructed image, stellar velocity and velocity
dispersion maps. From Emsellem, Fathi, Wozniak et al., in preparation.
[Note that the central 5~arcsec should be taken with caution as the 
stellar kinematics is strongly perturbed by wide and strong emission
lines].}
\label{fig:n1068}
\end{figure}

We have then conducted a small series of numerical N-body $+$ SPH
simulations using up to 2400000 and 30000 particles for the stellar
and gas components respectively. Starting from axisymmetric initial
conditions, the goal was to trigger the formation of a bar system
which would look like the one seen in the inner region of NGC~1068.
Tuning the initial conditions (proper mass distribution and dynamics), we have thus obtained
a time-evolving model the snapshots of which were
compared to the observed {\sc SAURON} maps. The first attempt at doing
this is presented in Fig~.\ref{fig:n1068} where the observed and
modeled stellar kinematics are presented. We could easily reproduce
the overall shape as well as the amplitudes of the stellar velocity
and dispersion fields.  Obviously the surface brightness image
reconstructed from the {\sc SAURON} datacube does not exhibit the
strong bar present in the numerical model as the former is
perturbed by strong stellar populations gradients, star
formation and emitting line regions. 

Apart from a detailed study of the kinematics of NGC~1068, 
the main goal is to now use these data and
models to quantify the gas fueling rate in the central few hundreds of
parsecs. Resonances in tumbling potential acts as boundaries where the orbital
structure can abruptly change. For $m=2$ modes such as bars and spirals,
the presence and strength of the Lindblad Resonances are critical in 
determining the dynamical characteristics of the system. Also, depending on 
the properties of the supporting medium (dissipative or collisionless), 
density waves are allowed (or not) to cross such barriers.
In this context, bars are efficient at driving gas toward their Inner
Linbdlad Resonance (if present). In the case of NGC~1068,
the location of the ILR has been estimated by Schinnerer et al.
(2000) to be at a radius of about 5~arcsec, or 350~pc. 
Gas indeed seems to be funneled towards this region, 
but the blending of numerous and complex gas emitting components
prevented us to investigate the fate of the gas further in.

\section{Bars in the Sombrero Galaxy and NGC~3115}

In order to probe the region inside the ILR, we should focus
onto less active objects. One galaxy hosting a weakly active nucleus, and a candidate for the
presence of a $\sim 10^9$~M$_{\odot}$ black hole (Emsellem et al. 1996
- note the different distance used in that paper - ,
Kormendy et al. 1996), is the Sa spiral M~104 (NGC~4594)
known as the Sombrero: its uncommon appearance is due to its high
inclination, its large bulge, and its prominent dust lanes at large
radii. This spiral galaxy exhibits strong evidence for
the presence of a large-scale bar as emphasized in two papers of a
series (Emsellem 1995; Emsellem et al. 1996). This argument was
initiated by the analysis of a resonance diagram (Emsellem et al.
1996, Fig.~22) which associated the different rings (CO, HI, ionized
gas, stellar) to different resonances (OLR, UHR) of a tumbling bar.
In this context, the ILR is situated at about 20~arcsec outside 
the bright inner secondary disk. 

Double disk structures such as the one in the Sombrero galaxy are
common in early-type spirals, and have been suggested to be formed via
bar driven evolution (see e.g. van den Bosch \& Emsellem 1997).  In
this scenario, the inner disk was mainly built from gas accretion
inside the ILR of the bar and subsequent star formation.  In the case
of the Sombrero galaxy, there is a spiral-like gaseous component
inside this inner disk, tentatively interpreted by Emsellem \& Ferruit
(2000) on the
basis of high resolution photometry and two-dimensional spectroscopy
to be the signature of a secondary bar driven structure.
The ILR of this secondary bar would thus be located at about 1~arcsec
or 47~parsec.  If confirmed, this would obviously have important
consequences on the estimate of the central black hole mass in this
galaxy. Although the Sombrero galaxy may be a rather unique object in the
nearby Universe, this would also have implications on our
understanding of the way gas can be funneled towards the nuclear region.

As emphasized above, early-type disk galaxies with double disks are
rather common, one of the best known example being the quiet S0 galaxy
NGC~3115. It exhibits a very thin and bright disk inside its
central 3~arcsec (Kormendy et al. 1996). The central value of its stellar velocity dispersion
is on the high side (about 450~km/s as measured with FOS), which has
been used to argue for the presence of a central dark mass (Kormendy
et al. 1996 and references therein). We have used both integral
field spectroscopy and high resolution FOS data to better constrain
the mass of the presumed black hole (Emsellem, Dejonghe, Bacon 1999):
two and three-integral models provided a range of acceptable masses
of $[4.5 - 13]$~$10^8$~M$_{\odot}$ with an overall best fit
three-integral model with 6.5~$10^8$~M$_{\odot}$ (note that the best
fit two-integral model has M$_{bh} = 9.4$~$10^8$~M$_{\odot}$).
\begin{figure}
\centering
\includegraphics[width=85mm]{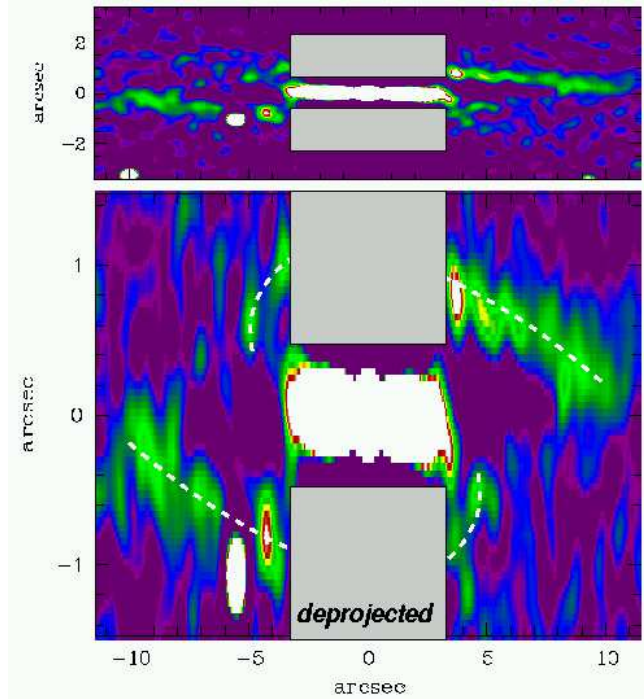}
\caption{Evidence for an inner stellar spiral in NGC~3115.  Top panel:
unsharp masking of a WFPC2/F555W ($V$ band) image of NGC~3115.
Bottom panel: same image but deprojected using an inclination of
$86\deg$ (Emsellem et al. 1999) and assuming the residual intensity
lies in the equatorial plane.  An hypothetic sketch for a
two-arm spiral is superimposed on the deprojected image.  High
frequency structures could not be derived in the grey areas due
to the presence of the bright inner disk.}
\label{fig:n3115}
\end{figure}

Is the bright and thin disk the consequence of bar-driven gas
accretion as sketched above, and is there a link between the central
massive black hole (if present) and this disk? Isophotes in the inner
region of NGC~3115 have varying shapes: they are (obviously)
disky in the central 3~arcsec, and become slightly boxy
when the luminosity profile of the inner disk drops. This is
qualitatively similar to what is observed in another edge-on S0,
NGC~4570, where the signature of a tumbling potential was
advocated. In NGC~3115, the boxiness is in fact partly due to the
presence of a weak centrally symmmetric structure resembling a
two-arms stellar spiral. This is illustrated in Fig.~\ref{fig:n3115}
where a simple unsharp masking of a WFPC2 image reveals this spiral,
which seems connected to the inner disk. Although a complete
stability analysis should be conducted before concluding too hastily,
such a weak (non self-gravitating) structure should require a driver, a
forcing tumbling term in the potential, namely a bar. This picture
would be consistent with the hypothesis that observed double disk
morphologies are linked to a bar driven secular evolution.
Again, this would be an important issue to take into account when
estimating the mass of the central dark object.  
And the question of whether or not the central dark object has grown
significantly during this evolutionary process would then remain.

\section{Remnant Signatures of Gas Fueling}

In our search for signatures of past or present gas fueling, we have
conducted a small study of the stellar kinematics of galaxies hosting
active galactic nuclei, mostly Seyfert~2 ones. This was originally
intended as a tool to look for possible links between the
gravitational potential and the activity of the nucleus. Our first
observations included {\sc ISAAC/VLT} near-infrared spectroscopy of 3
Seyfert~2 galaxies with photometrically detected double bars. We first
confirmed the expected kinematical decoupling of the inner structure.
A surprising result came from the stellar velocity dispersion
profiles, which, in all three cases, exhibits a significant drop at
the centre. These central dispersion drops (CDDs) could not be
explained via simple dynamical models which all predicted rising or
flat dispersion profiles. 
\begin{figure}
\centering
\includegraphics[width=100mm]{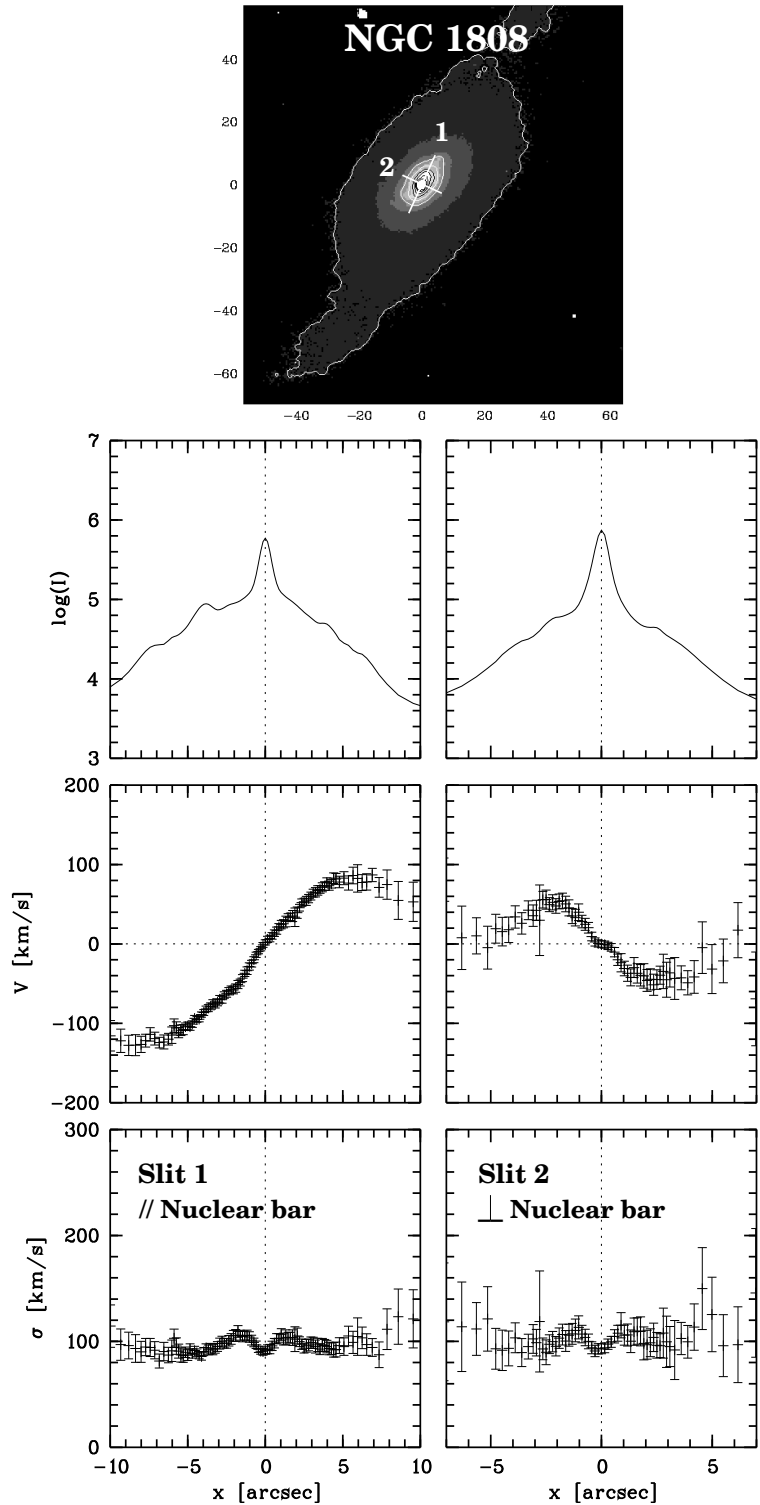}
\caption{{\sc ISAAC/VLT} stellar kinematics of NGC~1808 along the
major (left panels) and the minor (right panels) axis of the inner bar. 
An $H$ band image is presented at the top with a sketch of the extent
of the slits. The central dispersion drop is clearly visible along
both directions. Adapted from Emsellem et al. (2001).}
\label{fig:debca}
\end{figure}

We therefore suggested that these drops are transient features due to
the presence of a dynamically cold stellar component formed during an
episode of central gas accretion driven by a bar. This scenario
has to be confirmed with further observational and theoretical
evidences. However, we have already conducted a series of numerical
N-body $+$ SPH simulations, including star formation, which indeed
show that these CDDs naturally appear when gas is driven towards
the centre by a bar and a new stellar disk of young stars forms
(Wozniak, Combes, Emsellem, Friedli, in preparation).
We have also looked more systematically for CDDs in published data.
Although most of the stellar kinematics available in the literature is
of medium quality and resolution, we have found a few
instances of such drops, almost always in active, barred galaxies.
We just obtained new spectroscopic observations in the near-infrared,
where dust extinction is a less severe problem: we should be soon able
to test the link between the activity of the nucleus (AGN or
starburst) and the inner potential (barred or not). We provide here a
beautiful example of a CDD discovered in the barred galaxy NGC~3623 in
the course of the {\sc SAURON} survey (see Bacon et al. 2001, de Zeeuw et al.
2002): the drop in the dispersion is clear (although there is some
extinction due to dust), and is coincident with a cold component
detected in the stellar velocity field.
\begin{figure}
\centering
\includegraphics[width=100mm]{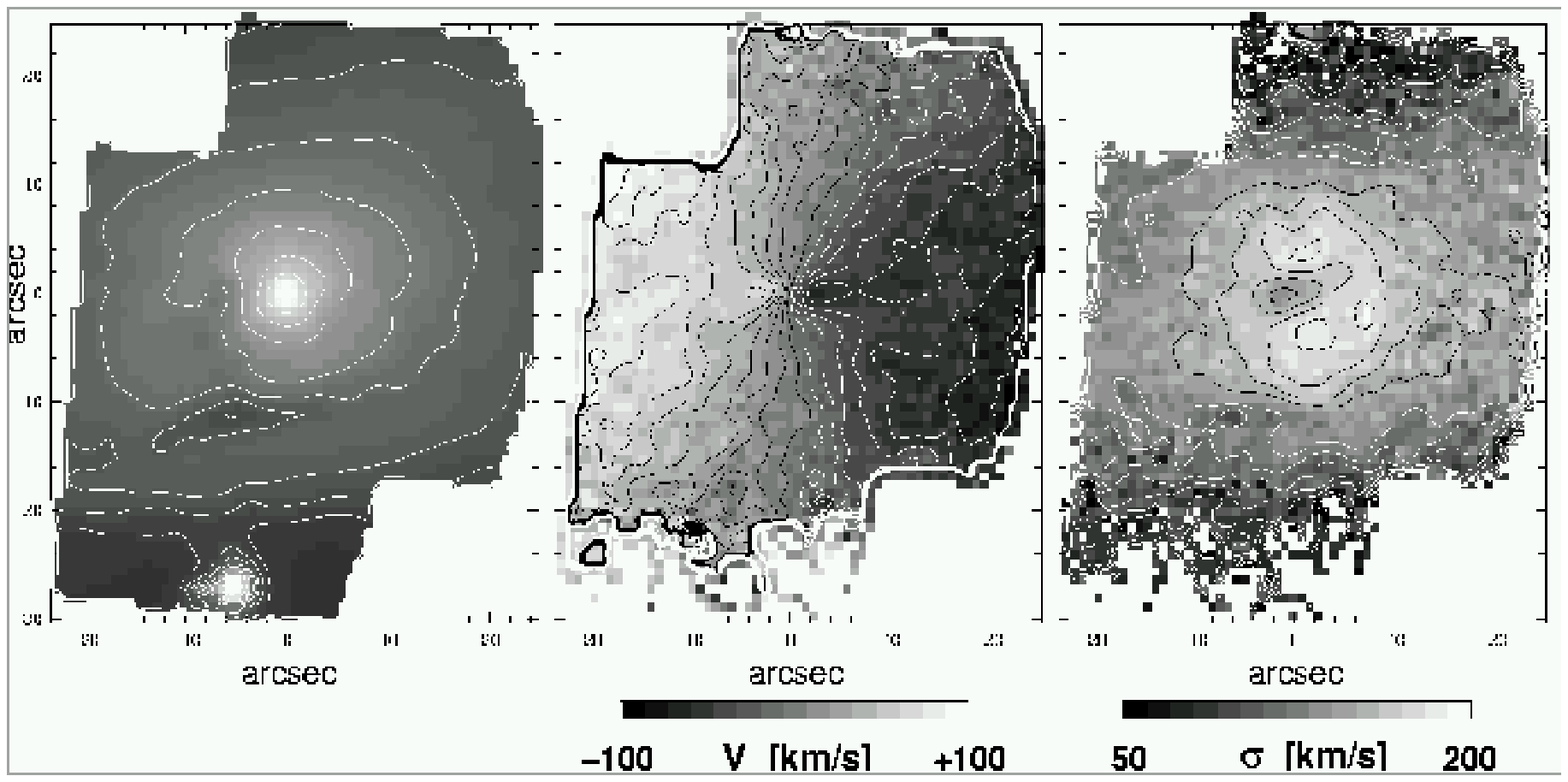}
\caption{Two-dimensional stellar kinematics of NGC~3623 obtained with the {\sc
	SAURON} spectrograph mounted at the WHT. From left to right:
		reconstructed surface brightness image, stellar velocity and
		stellar velocity maps. Note the drop in the dispersion as well
		as the clear signature of a disk in the velocity map.}
\label{fig:n3623}
\end{figure}

\section{Lopsided Density Distributions}

I should finally say a word about the so-called $m=1$ modes,
associated with lopsided density distributions (see Combes 2001,
Emsellem 2002). I will focus on the very striking case of
M~31 which has been discussed as a candidate for the
presence of a supermassive black hole since almost 15 years
(Kormendy 1988; Dressler \& Richstone 1988). The nucleus of M~31
has been known to harbour a double peaked structure since the
Stratoscope~II images (Light et al.  1974), and has been studied in
detail with WFPC/HST images by Lauer et al. (1993). Many papers
have been published since, attempting to build a physical picture
for this puzzling morphology. 

\begin{figure}
\centering
\includegraphics[width=105mm]{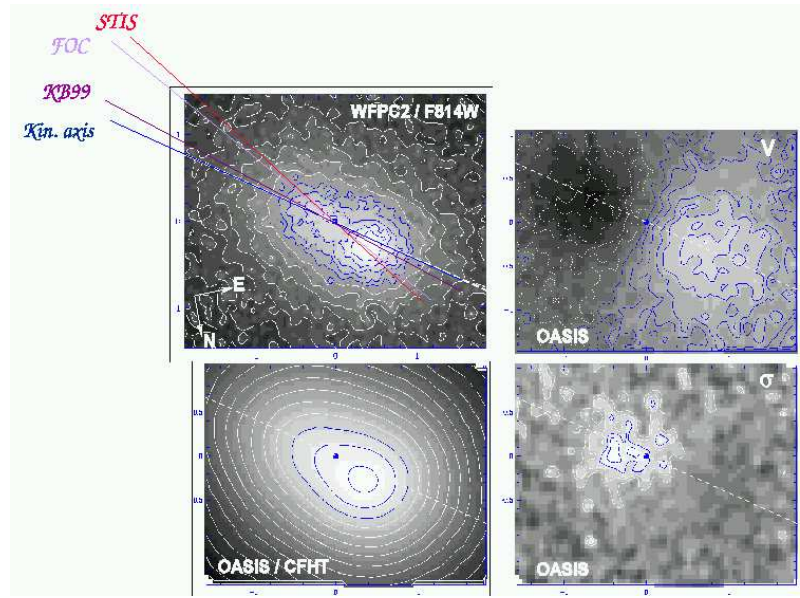}
\caption{The nucleus of M~31. The top left panel shows a WFPC2/HST
	image with overimposed isophotes, and a sketch of some long-slit
		used to probe the kinematics of this nucleus. The three other
		panels show {\sc OASIS} integral field spectroscopy of M31's
		nucleus: (bottom left) reconstructed image, (top right) stellar
velocity map, (bottom right) stellar velocity dispersion map. Adapted
from data presented in Bacon et al. (2001).}
\label{fig:m31}
\end{figure}

Two-dimensional kinematics were obtained by Bacon et al. (1994),
showing a clear offset of the dispersion maximum with respect to the
centre of the outer isophotes (P2), already hinted in the previous
long-slit works. Bacon et al. (1994) also built simple dynamical Jeans
(axisymmetric) models estimating the mass of the central dark object
to be $\sim 7\; 10^7$~M$_{\odot}$. The first serious attempt at
modeling the lopsided luminosity distribution was conducted by
Tremaine (1995) who proposed to explain the offset luminosity (P1) and
dispersion peaks by a set of aligned eccentric orbits around a
dominating massive black hole. The hypothesis of a decaying cluster
destroyed via the tidal forces of the black hole has also been
examined (Emsellem \& Combes 1997): although it provided good
fits to the existing data with the first self-consistent models for
this nucleus (black hole masses between 7 and
10~$10^7$~M$_{\odot}$), the overall lifetime of the secondary peak
was too short lived for this scenario to be a viable alternative. This
hypothesis was fully rejected when detailed spectroscopic analysis demonstrated
that, besides a UV excess at the presumed location of the black hole
(P2), the stellar population within the nucleus was rather homogeneous
(Kormendy \& Bender 1999). Subsequent high resolution spectroscopy
(Statler et al. 1999), and modeling efforts (e.g. Statler 1999;
Salow \& Statler 2001; Jacobs \& Sellwod 2001) further constrained the detailed dynamical
structure of this nucleus. Then Bacon et al. (2001) reexamined most of the
available data, also adding new {\sc OASIS} integral field spectroscopy, 
to reconcile the different photometric and kinematical profiles. As
shown in Fig.~\ref{fig:m31}, only two-dimensional spectroscopy can
really probe the complex structure of this nucleus. 

Bacon et al. (2001) also built numerical self-consistent models to fit the photometry as 
well as the stellar velocity profile. We found that indeed a slowly
rotation prograde $m=1$ mode could reproduce the observed photometry.
Such a mode exists when the gravitational potential is nearly
keplerian (in the limit of which the epicyclic frequencies are equal
to the circular frequency). However, the central dark mass should not
fully dominate the potential, to allow the self-gravity to act and
help the mode to grow. In the case of M~31, the mass of the disk
should thus be between 20 and 40\% of the mass of the central black
hole, consistent with the advocated value of 7~$10^7$~M$_{\odot}$.
As suggested in Emsellem (2002a,b), the nucleus of M~31 is probably
the result of gas accretion, which may be related to the present overall 
structure of the galaxy (bar, inner dust lanes).

\section{Conclusions}

I hope this eclectic selection of illustrations helps to demonstrate
that density waves are indeed present in the central regions of
galaxies, and that there is probably a link between these waves and the central
dark massive object. This link runs in both directions: inner density
waves can help bringing gas in the central 10 or 50 parsecs, closer
to seed black holes which may ultimately grow, and central dark masses
can strongly influence the dynamical state of this very same region.
In this context, bars and $m=1$ (non-keplerian) may play a significant
role.  Signatures for such fueling can be found in the stellar
kinematics, with the so-called CDDs being presumably rather common in disk galaxies.
Finally, we should keep all this in mind, when modeling the dynamics
of nearby galaxies in our search for supermassive black holes.

\end{document}